\documentclass[dvipsnames]{osa-article}

\journal{oe}

\usepackage{xcolor}
\usepackage{graphicx}
\usepackage{bm}
\usepackage{siunitx}
\usepackage[title]{appendix}
\usepackage[textsize=scriptsize]{todonotes}
\usepackage{xargs}
\newcommandx{\vnm}[2][1=]{\todo[color=blue!30, #1]{vnm: #2}}
\newcommandx{\bdl}[2][1=]{\todo[color=Emerald!30, #1]{bdl: #2}}
\newcommandx{\csm}[2][1=]{\todo[color=red!30, #1]{csm: #2}}

\begin{document}

\title{Improving holographic particle characterization by modeling spherical aberration}


\author{Caroline Martin,\authormark{1}, Brian Leahy,\authormark{1} and
Vinothan N. Manoharan,\authormark{1,2,*}}

\address{
\authormark{1}Harvard John A. Paulson School of Engineering and Applied
Sciences, Harvard University, Cambridge MA 02138, USA\\
\authormark{2}Department of Physics, Harvard University, Cambridge MA 02138, USA\\
}

\email{\authormark{*}vnm@seas.harvard.edu}

\begin{abstract}
  Holographic microscopy combined with forward modeling and inference
  allows colloidal particles to be characterized and tracked in three
  dimensions with high precision. However, current models ignore the
  effects of optical aberrations on hologram formation. We investigate
  the effects of spherical aberration on the structure of
  single-particle holograms and on the accuracy of particle
  characterization. We find that in a typical experimental setup,
  spherical aberration can result in systematic shifts of about
  \SI{2}{\percent} in the inferred refractive index and radius. We show
  that fitting with a model that accounts for spherical aberration
  decreases this aberration-dependent error by a factor of two or more,
  even when the level of spherical aberration in the optical train is
  unknown. With the new generative model, the inferred parameters are
  consistent across different levels of aberration, making particle
  characterization more robust.
\end{abstract}

\renewcommand{\vec}[1]{\mathbf{#1}}
\newcommand{\unit}[1]{\mathbf{\hat{#1}}}
\newcommand{\tensor}[1]{\bm{#1}}

\renewcommand{\Im}{\mathrm{Im}}
\renewcommand{\Re}{\mathrm{Re}}

\newcommand{\hologram}{H}

\newcommand{\micron}{\SI{}{\micro\meter}}
\newcommand{\microliter}{\SI{}{\micro\liter}}
\newcommand{\phip}{\varphi_\mathrm{p}}
\newcommand{\zp}{z_\mathrm{p}}
\newcommand{\rhop}{\rho_\mathrm{p}}

\newcommand{\thetao}{\theta}

\newcommand{\wavefront}{\Phi}

\newcommand{\Edet}{\vec{E}_\mathrm{sc}}
\newcommand{\Eindet}{\vec{E}_\mathrm{in}}

\newcommand{\farfield}{\tensor{S}}

\newcommand{\Izero}{\mathcal{I}_0}
\newcommand{\Itwo}{\mathcal{I}_2}

\newcommand{\lensangle}{\beta}

\newcommand{\crosssection}{\Sigma}

\section{Introduction}
The combination of digital holographic microscopy, forward modeling, and
statistical inference allows colloidal particles to be characterized and
tracked with high precision over a large depth of
field~\cite{alexander_precise_2020}. In contrast to the traditional method of
 interpreting holograms by reconstruction~\cite{gabor_new_1948, schnars_digital_2002},
 where the fields scattered by the objected are recovered by physically or numerically
 shining light through the recorded hologram, in forward-modeling
approaches a scattering theory is directly fit to a minimally
processed hologram~\cite{ovryn_imaging_2000, lee_characterizing_2007}. This
approach yields estimates of the
particle's three-dimensional position, index of refraction, and size.
The forward-modeling approach has the advantage that
 the position and properties of the object can be inferred directly
 from the hologram, whereas reconstructions must be further processed
to recover this information. When the objects are spheres
approximately as large as the wavelength, their reconstructions
often do not resemble spheres~\cite{pu_intrinsic_2003}, making it difficult to 
precisely extract their positions from the reconstruction. Furthermore, the
forward-modeling approach yields estimates of the particle's index
of refraction, which cannot be directly inferred from a
reconstruction. The precision of the quantities inferred from
forward-modeling and fitting, combined with
the high acquisition speed of the holographic microscope, make the
approach useful for many applications, including
microrheology~\cite{cheong_holographic_2009}, visualizing the dynamics
of colloidal clusters~\cite{fung_holographic_2013}, and studying
bacterial swimming~\cite{wang_tracking_2016}.

However, the most commonly used forward models ignore optical
aberrations, which exist in all imaging systems. Not
modeling such aberrations can lead to systematic errors in the
inferred parameters. Although many
techniques have been demonstrated to correct for the effects of
aberrations and artifacts on reconstructions~\cite{stadelmaier_compensation_2000,
  grilli_whole_2001, ferraro_compensation_2003, colomb_automatic_2006,
  nicola_recovering_2005, nguyen_automatic_2017, verrier_image_2014,
  denis_twin_2005}, there has been little
work on accounting for these effects in forward models. 
Recent models do include the phase effects introduced by an
ideal objective lens~\cite{leahy_large_2020, alexander_precise_2020},
but they do not include aberrations in either the objective or
experimental setup. Moreover, the effects of aberrations on the accuracy
of parameters inferred through model-based approaches have not yet been
examined.

\begin{figure}
  \centering
  \includegraphics[width=0.75\textwidth]{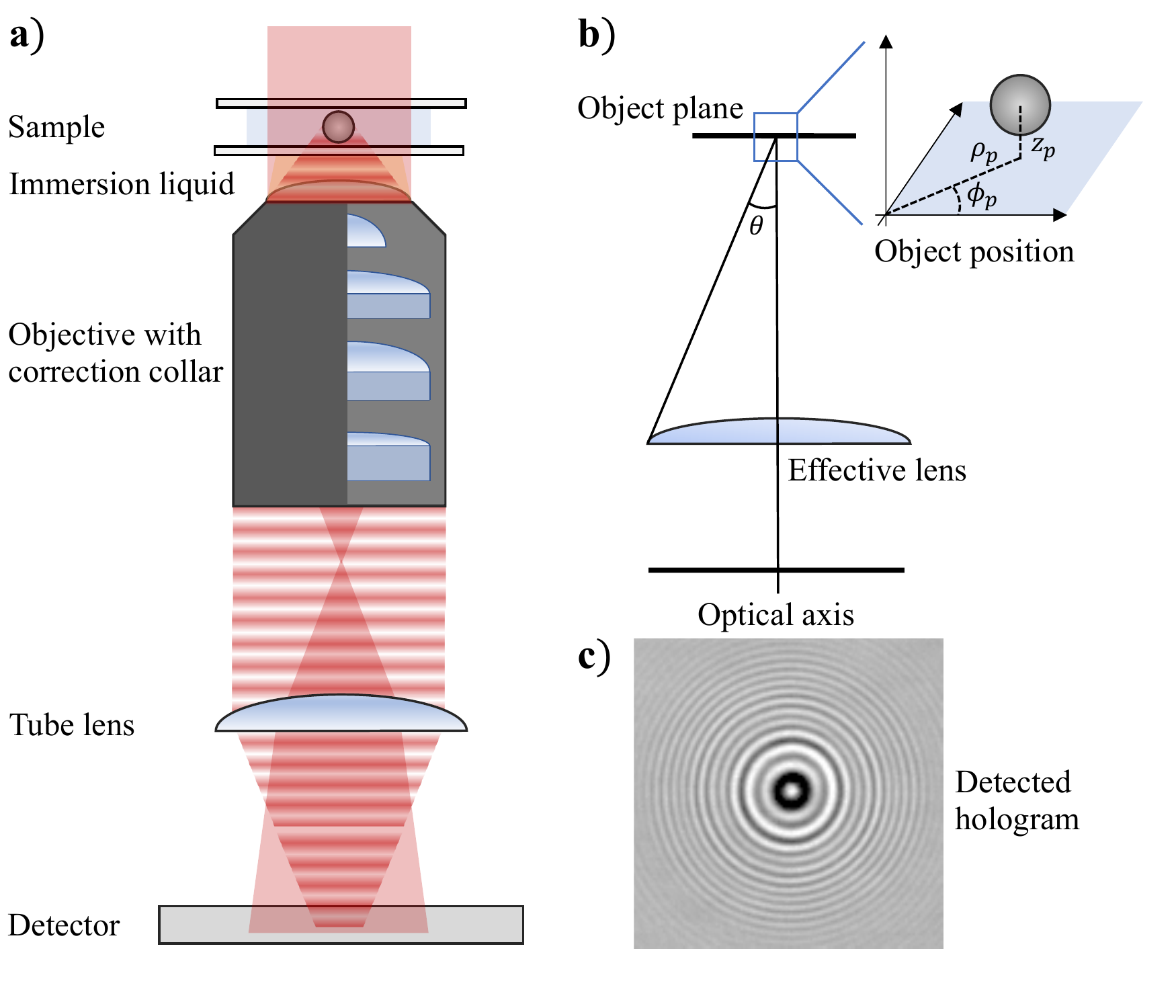}
  \caption{(a) Diagram of the optical train in a typical in-line digital
    holographic microscope. Collimated coherent light (red) illuminates
    a sample chamber consisting of an object in a medium
    between a glass slide and coverslip. Here, we show an immersion
    objective with a correction collar. (b) We treat the optical train
    as a single effective lens pupil with polar angle $\thetao{}$, as
    defined from the optical axis. The object's position in the object
    plane is defined by $(\rhop{}, \phip{}, \zp{})$ in cylindrical
    coordinates.
    (c) A measured hologram from a \SI{1.05}{\micro\meter}
    polystyrene sphere sitting \SI{7.5}{\micro\meter} above the focal
    plane, illuminated with \SI{660}{\nano\meter} light. We record the
    hologram with a water-immersion lens with a numerical aperture of
    1.20 and set the correction collar to minimize aberration.}
  \label{fig:hologram}
\end{figure}

It is important to understand the effects of aberrations because most
microscopy experiments are subject to them. Here, we focus on spherical
aberration. Although most high-quality objectives are corrected for
spherical aberration near the focus, holographic microscopy is often used to
image particles far from the focal plane, where the aberration may be
less well corrected. Furthermore, the interface between a liquid sample
and a glass coverslip can lead to spherical aberration when the
refractive index of the objective immersion fluid is not matched to that
of the sample medium~\cite{hell_aberrations_1993}. Water-immersion
objectives are therefore used to minimize aberrations when aqueous
samples are imaged. But even with a water-immersion objective, spherical
aberration can be introduced by the coverslip interface. This aberration
can be corrected by setting the objective's correction collar, which
adjusts the position of a movable central lens group within the
objective (Figure~\ref{fig:hologram}a), to compensate for the thickness
 of the coverslip, which is typically \SIrange{0.10}{0.20}{\mm}. If the
correction collar is set incorrectly, however, spherical aberration is
induced. Thus, unless the microscopist carefully measures the thickness
of each coverslip---a laborious task when many samples must be
imaged---the holograms will likely be subject to spherical aberration.
To maximize the precision of tracking and characterization, we must
therefore consider the effects of aberrations that can be induced by
the experimental setup, as well as those inherent to the objective.

In this article, we experimentally examine the effects of spherical
aberration on holograms captured in an in-line holographic microscope
and develop a model to describe these aberrations. We
fit this model directly to aberrated holograms and extract
information about the particle as well as the aberrations. This
method does not require reconstruction or image processing.
We find that modeling the effect of aberrations
 improves agreement between the model predictions and the 
 experimental data, leading to lower residuals
between the best-fit holograms and data across levels of induced
aberration. We also find that the inferred parameters do not change as
the level of aberration in the optical train increases. Modeling the
aberrations therefore allows one to more accurately characterize
colloidal spheres even if one does not know whether the experimental
setup is spherically aberrated, or by how much. The robustness of the
fits with this new forward model to aberrations, either induced in or
inherent to the optical system, could simplify experiments and reduce
systematic errors.

\section{Effect of spherical aberration on hologram structure}
To explore the effects of typical levels of spherical aberration on
hologram structure, we image an immobilized polystyrene microsphere with
a reported radius of \SI{1.05}{\micro\meter} (density
\SI{1.055}{\gram/\milli\liter}, index 1.591 at \SI{590}{\nano\meter},
Invitrogen S37500) under varying levels of spherical aberration. To
immobilize the microsphere, we fill an inverted sample chamber with a
\SI{0.001}{\percent} w/v colloidal suspension in a \SI{0.2}{\milli M}
NaCl aqueous solution. The particles then sediment and stick to the
slide, likely because the salt screens any electrostatic repulsion
between the sphere and glass. We then reorient the sample chamber. The
particles remain immobilized on the top slide, approximately
\SI{150}{\micro\meter} above the interface between the bottom slide and
immersion lens.

The particles are illuminated with a diode laser (\SI{660}{\nano\meter}
wavelength, \SI{120}{\milli\watt} power, Opnext HL6545MG) driven at
\SI{130}{\milli\ampere} (Thorlabs LDC205C). We use a Nikon Eclipse Ti
TE2000 microscope with a water-immersion objective and correction collar
(Plan Apo VC 60$\times$/1.20 WI, Nikon, \SI{300}{\micro\meter} 
working distance) and a $1024 \times 1024$-pixel
CMOS monochrome sensor array (PhotonFocus A1024) to capture the
hologram. Before analyzing each hologram, we subtract the average of 
dark-count images taken without illumination and divide the resulting 
hologram by averaged background images taken with no particles in 
the field of view. This image processing differs from the
double-exposure method of reconstruction-based holography, which
removes phase shifts and thus aberrations in the resulting
phase-contrast image. We do not attempt to reconstruct the background
to obtain a phase map~\cite{ferraro_compensation_2003}. 
Instead, our processing method removes
stray illumination in the optical train and accounts for non-uniform
illumination and artifacts.

We adjust the level of induced spherical aberration by varying the
correction collar setting on the objective. The correction collar on our
objective corrects for coverslip thicknesses of
\SIrange{0.13}{0.19}{\milli\meter}. Setting the collar to the thickness
of the coverslip, measured at \SI{0.17}{\milli\meter}, minimizes
spherical aberration; setting it to the furthest available setting,
\SI{0.13}{\milli\meter}, maximizes spherical aberration. At the
maximally aberrated setting, we expect aberration in the phase from
\SI{0.04}{\milli\meter} of glass, which is the difference between the
measured and corrected thicknesses. This difference should induce a
phase shift $\Delta \phi$ between the paraxial and off-axis rays. For a
homogeneous material with refractive index $n_2$ and thickness $h$
embedded in a medium with refractive index $n_1$, $\Delta \phi$ is given
by
\begin{align}
\Delta \phi= h k \left(n_2 - n_1\right) \left[2 \frac{n_1}{n_2}
\sin^2 \left(\frac{\theta}{2}\right) + 2 \left(n_2 + n_1\right) \frac{n_1^2}{n_2^3}
\sin^4\left(\frac{\theta}{2}\right) + \ldots \right],
\label{eqn:aberration}
\end{align}
where $k$ is the wavenumber and $\theta$ is the angle of incidence of
the off-axis ray~\cite{sheppard_aberrating_1991}. For a glass layer
($n_2=1.515$) with thickness $h=\SI{0.04}{\milli\meter}$ surrounded by
water ($n_1=1.33$), and at the maximum angle of incidence set by the
numerical aperture of the lens ($\mathrm{NA}=1.2$), we calculate a
maximum phase shift corresponding to approximately 52 wavelengths. 
In the above expression, the $\sin^2 (\theta/2)$ term corresponds to defocus, 
which accounts for 35 wavelengths of phase shift, and the $\sin^4 (\theta /
2)$ term corresponds to the primary spherical aberration, which accounts for the
remaining 17 wavelengths. The higher order terms correspond  to higher order 
spherical aberration.  To minimize the aberration induced by the
coverslip, we correctly set the correction collar to the precisely
measured thickness of the coverslip (\SI{0.17}{\milli\meter}). In this
case, we expect any remaining aberration to be intrinsic to the optical
system, rather than induced by the experimental setup. 
We note that because high-numerical-aperture
objectives are designed to image objects in the focal plane, they
may not be well corrected for aberrations far from the focus.

\begin{figure}
  \centering
  \includegraphics{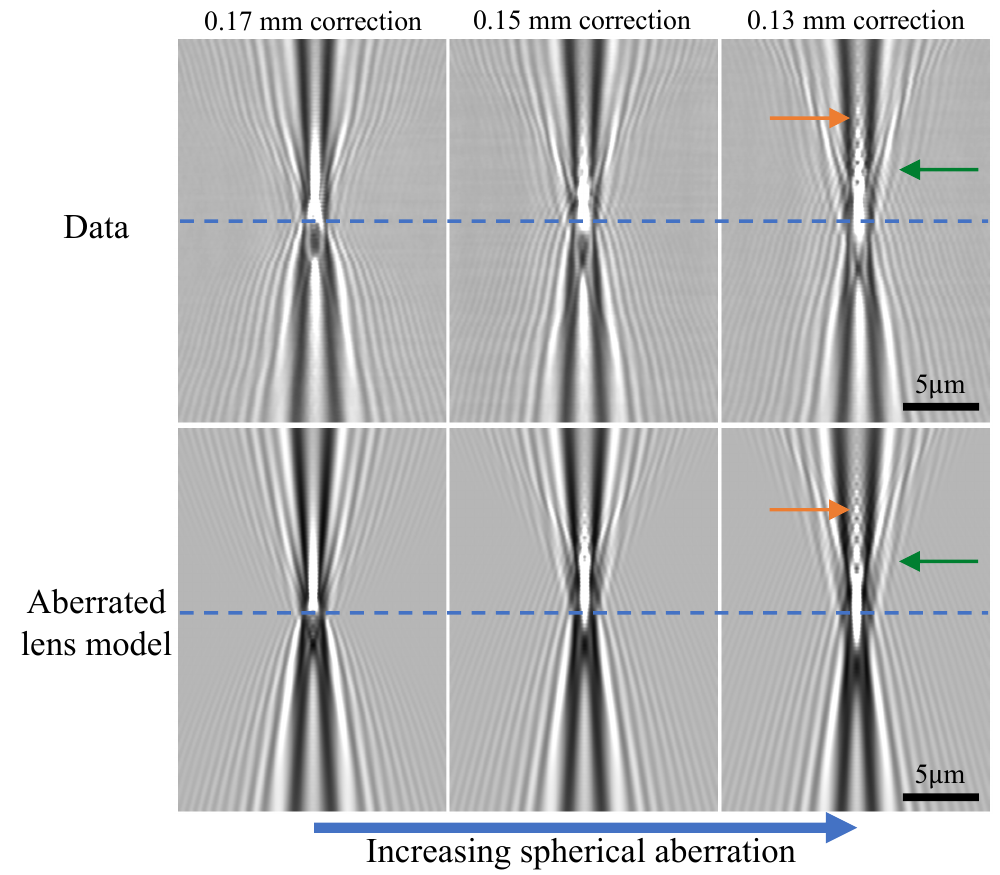}
  \caption{$x$-$z$ cross-sections of stacks of holograms of an
    immobilized \SI{1.05}{\micro\meter}-radius polystyrene microsphere,
    taken from \SI{15}{\micro\meter} below to \SI{15}{\micro\meter}
    above the focus in \SI{0.25}{\micro\meter} intervals at three
    correction collar settings. Adjusting the correction collar induces
    spherical aberration, which results in on-axis oscillations in
    brightness (orange arrow), overlapping fringes close to the focal
    plane (green arrow), and defocus. The blue line marks the focal
    plane of the least aberrated system; all stacks are taken with the
    lens in the same physical position. The top row of cross-sections
    shows experimental data. The bottom row shows calculations from a
    model that includes spherical aberration. By changing $a_0$ in
    the aberration series (primarily fourth order in the phase
    deviation), we are able to produce cross-sections that capture the
    distortions in the experimentally observed holograms.}
  \label{fig:x-z}
\end{figure}

To examine how increasing the aberration affects the hologram structure,
we record ``stacks'' of holograms of the immobilized sphere by sweeping
the focus through the particle at three different correction-collar
settings. We then examine the $x$-$z$ cross-sections of these hologram
stacks, where each section shows the intensity of the hologram through
the central fringe (Figure~\ref{fig:x-z}). The particle height
determines the spacing of the fringes in the resulting hologram, with
the fringe spacing increasing with increasing distance from the focal
plane. The $x$-$z$ cross-sections therefore have a cone-like structure,
with a bright center above the focus and a dark center below. With no
spherical aberration, we expect a single focal point at which the center
of the hologram transitions from bright to dark, which we observe in the
cross-section at \SI{0.17}{\milli\meter} correction.

As we change the correction collar setting, the increase in aberration
introduces several noticeable changes in the recorded hologram
structure. First, oscillations between bright and dark points appear
along the central axis (orange arrows in Figure~\ref{fig:x-z}), and the
number of oscillations increases as we increase the aberration. Second,
we observe an overlap of fringes near the focus. Instead of converging
at the focal plane, the bottom cone structure converges at a point above
the focal plane with increasing aberration, resulting in distortions due
to the overlap of the fringes of the top and bottom cone near the focal
plane (green arrows in Figure~\ref{fig:x-z}). Finally, the position of
the focal plane shifts as the aberration increases, as highlighted in
Figure~\ref{fig:x-z} by the increasing distance of the focal planes from
the blue dotted line.

These changes in structure are the result of the angle-dependent
phase shift due to increasing spherical aberration. Spherical
aberration changes the phase of the off-axis rays. The interference
between these peripheral rays and the paraxial rays produces the bright
and dark points that we see in the data; the up-down asymmetry is
characteristic of spherical aberration~\cite{born1983principles}. The
overlap of fringes as the cone structures do not come to a single point
is also indicative of spherical aberration, as there is no longer a
single well-defined focal plane in the presence of these aberrations.
Finally, the observed defocus effect is also characteristic of
aberration induced by index mismatch, as described above.

\section{Modeling the effect of spherical aberration}
These results show that spherical aberration caused by small offsets in
the correction collar can induce significant changes in the recorded
holograms across a wide depth of field. To infer accurate particle
parameters under typical levels of spherical aberration, we need a
forward model that accounts for these effects.

To model the effects of aberration on a recorded hologram, we build upon
the treatment for modeling an unaberrated lens from Leahy, Alexander,
and coworkers~\cite{leahy_large_2020}. In an unaberrated system, all
rays emitted from a point source on the focal plane come to a focus at a
single point on the detector. By Fermat's principle, these rays traverse
the same optical path length through the imaging system. In an aberrated
system, the rays do not come to a focus at a single point, and therefore
the phase of each ray differs. These phase aberrations can be quantified
by the deviation $\wavefront{}$ of each ray's phase from its ideal
value, as measured on the lens pupil. In the presence of general
aberrations, $\wavefront{}$ depends both on the position on the lens
pupil and on the source's in-plane position~\cite{born1983principles}.
For spherical aberration, however, $\wavefront{}$ is a function of the
polar angle $\thetao{}$ only, defined in Figure~\ref{fig:hologram}b. For
incident light polarized along the $x$-direction $\unit{x}$ of the
detector plane, propagating the fields through the optical train yields
the incident ($\Eindet{}$) and scattered fields ($\Edet{}$) on the
detector:
\begin{equation}
  \Eindet{} \propto - {E_0} e^{i\wavefront{}(0)} \unit{x}
\end{equation}
\begin{equation}
  \begin{split}
    \Edet{} (\rhop{}, \phip{}, \zp{}) \propto \frac {E_0}{2} \Bigl\{ \bigl[& \Izero{}(k\rhop{}, k\zp{}) + \Itwo{}(k\rhop{}, k\zp{}) \cos(2\phip{}) \bigr] \unit{x}  \\
   &+\Itwo{}(k\rhop{}, k\zp{}) \sin(2\phip{}) \unit{y} \Bigr\},
  \end{split}
\end{equation}
where $(\rhop{}, \phip{}, \zp{})$ is the position of the particle in
cylindrical coordinates, as shown in Figure~\ref{fig:hologram}b, and
$\unit{y}$ is the unit vector along the $y$ direction. We have omitted
phase and amplitude factors common to both the incident and scattered
fields. In the presence of aberrations, the integrals $\Izero{}$ and
$\Itwo{}$ are defined as
\begin{align}
\Izero{} (u, v) = & \int_0^\lensangle
    \left[ S_\perp(\thetao{}) + S_\parallel(\thetao{})\right] J_0(u \sin \thetao{})
    e^{iv(1- \cos \thetao{}) } e^{i\wavefront{}(\thetao{})}
    \sqrt{\cos \thetao{}} \sin \thetao{} \, d\thetao{}
\label{eqn:integral_I0_def}
\\
\Itwo{} (u, v) = & \int_0^\lensangle
    \left[ S_\perp(\thetao{}) - S_\parallel(\thetao{})\right] J_2(u \sin \thetao{})
    e^{iv(1- \cos \thetao{}) }  e^{i\wavefront{}(\thetao{})}
    \sqrt{\cos \thetao{}} \sin \thetao{} \, d\thetao{},
\label{eqn:integral_I2_def}
\end{align}
where $J_n$ is the Bessel function of the first kind of order $n$, 
$S_\perp$ and $S_\parallel$ are the components of the far-field 
scattering matrix given by Mie theory, and
$\lensangle{}$ is the acceptance angle of the lens. 

In the unaberrated-lens model~\cite{leahy_large_2020}, $\wavefront{}$ is
constant and $\Izero{}$ and $\Itwo{}$ are changed only by a phase factor
$e^{i\wavefront{}}$ imparted by the lens. In the presence of spherical
aberrations, $\wavefront{}$ is a general function of $\thetao{}^2$, and
therefore $\wavefront{}$ is generally Taylor-expanded as an even
polynomial in $\thetao{}$. The constant term in this polynomial
corresponds to piston, which is irrelevant for in-line holographic
microscopy because it affects the incident and scattered fields
identically and therefore does not alter their interference. The
quadratic term corresponds to defocus, which is degenerate with $\zp{}$.
Thus, we represent $\wavefront{}(\thetao{})$ as a function $(1 - \cos
\thetao{})^2 \left[ a_0 + a_1 P_1 (1 - \cos\thetao{}) + a_2 P_2(1 - \cos
  \thetao{}) \right]$, where $P_\ell$ are Legendre polynomials and the
coefficients $a_\ell$ are parameters that describe the level of
aberration. Since $1 - \cos \thetao{} = \thetao{}^2 / 2 - \thetao{}^4 /
24 + \ldots$, this parameterization excludes piston and defocus
aberrations. The coefficients $a_0$, $a_1$, and $a_2$ allow us to
account for fourth- through eighth-order aberrations in the phase (or
third- to seventh-order in the ray displacements). While the coefficient
$a_0$ primarily describes fourth-order spherical aberration, the
coefficients $a_1$ and $a_2$ describe a mix of fourth-, sixth-, and
eighth-order aberrations. We choose to parameterize $\Phi$ in terms of
Legendre polynomials to make it easier to fit the aberration
coefficients to data; a parameterization in terms of ordinary
polynomials leads to stronger covariances between the inferred expansion
coefficients.

With this model for spherical aberrations, we are able to generate
holograms that capture the aberrated structure we observe in the data.
We generate $x$-$z$ cross-sections at evenly spaced intervals as $\zp{}$
moves through the focal plane (Figure~\ref{fig:x-z}, bottom), using
parameters based on the manufacturer's specifications of the microsphere
imaged in the top row. When the aberration coefficients are set to zero,
we recover the expected unaberrated structure: a double cone that
comes to a single focus where the hologram center transitions from
bright to dark. When we increase the aberrations by increasing $a_0$
(corresponding to primarily fourth-order phase aberrations), we
reproduce the effects we observe in the aberrated data, including the
on-axis oscillations in brightness and the overlapping fringe pattern
near the focal plane. Because our parameterization of the aberration
function separates defocus from spherical aberrations, we also adjust
$\zp{}$ in the generated holograms to account for the defocus in the
data.

The model does not capture all of the structure seen in the
cross-sections, particularly in the regions near the focal plane. For
example, the model does not accurately reproduce the intensity and
structure of the dark region below the focus. These discrepancies
suggest the need to model other effects, including perhaps other types
of aberrations. Nonetheless, our results show that accounting for
spherical aberration captures many of the aberration-dependent
distortions in hologram structure that can arise under typical
experimental conditions.

\section{Effect of spherical aberration on particle characterization}
To determine how spherical aberrations affect the accuracy of particle
characterization, we fit holograms of a single, immobilized
\SI{1.05}{\micro\meter}-radius polystyrene sphere at varying coverslip
corrections, recorded from \SI{30}{\micro\meter} above to
\SI{30}{\micro\meter} below the focus in \SI{0.25}{\micro\meter}
intervals. We fit both the aberrated-lens model described above and the
unaberrated-lens model~\cite{leahy_large_2020} (hereafter called the
``lens model'') to the measured holograms to infer the particle radius,
refractive index, and position, as well as the objective acceptance
angle $\beta$ and the field rescaling parameter $\alpha$. For the
aberrated-lens model, we also fit three coefficients of the aberration
series, $a_0$, $a_1$, and $a_2$ (see Appendix~\ref{appendix:methods} for
details on how we choose the order of the series required to describe
aberrations in the data). We use a Bayesian framework, as described in
Appendix~\ref{appendix:methods}. To avoid local minima in the posterior
probability density, we use a combination of nonlinear least-squares
fitting, covariance matrix adaptation evolution strategy, and
parallel-tempered affine-invariant Markov-chain Monte Carlo sampling.

\begin{figure}
  \centering
  \includegraphics{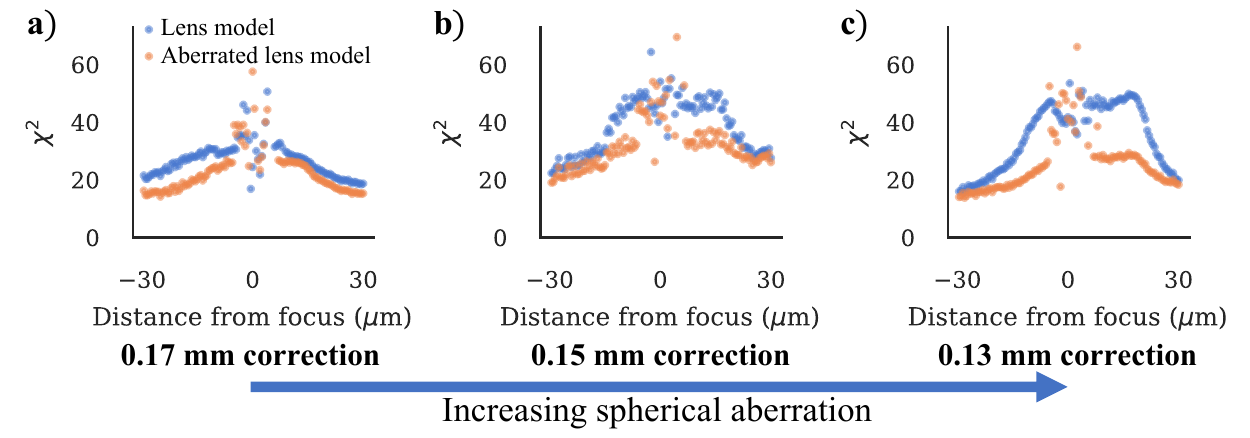}
  \caption{(a-c) The sum of squared residuals between holograms of an
    immobilized microsphere and best-fit holograms found with both the
    lens model and aberrated-lens model as a function of distance from
    focal plane and level of spherical aberration. While both models
    produce high residuals near the focal plane, outside this region,
    the aberrated-lens model produces consistently lower residuals than
    the lens model does. This improvement in goodness-of-fit occurs
    across $\zp{}$ and aberration level, with the largest improvement
    seen in the most aberrated system.}
  \label{fig:residuals}
\end{figure}

We find that the aberrated-lens model produces consistently better fits
to the data than the lens model, as measured by the sum of squared
residuals $\chi^2$ (Figure~\ref{fig:residuals}). When the particle is
within \SI{5}{\micro\meter} of the focal plane, fitting becomes
inconsistent for both models, and the residuals between the data and
best-fit holograms are large. When the particle is \SI{30}{\micro\meter}
from the focal plane, both models produce fits with comparable
residuals. As the microsphere approaches the focus, however, the
residuals found by the lens model increase significantly. Furthermore,
the lens model fits the data increasingly poorly as the aberration
increases. By contrast, the residuals found by fitting the
aberrated-lens model do not increase as sharply with decreasing distance
between the microsphere and focal plane and do not change as much with
the level of aberration. The aberrated-lens model also produces
consistently lower residuals than the lens model, with the largest
improvement in goodness-of-fit observed for the most aberrated system.
When the particle is farther than \SI{5}{\micro\meter} from the focal
plane, we find that modeling the effect of aberrations produces more
consistent residuals and improves goodness-of-fit at all aberration
levels.

\begin{figure}
  \centering
  \includegraphics{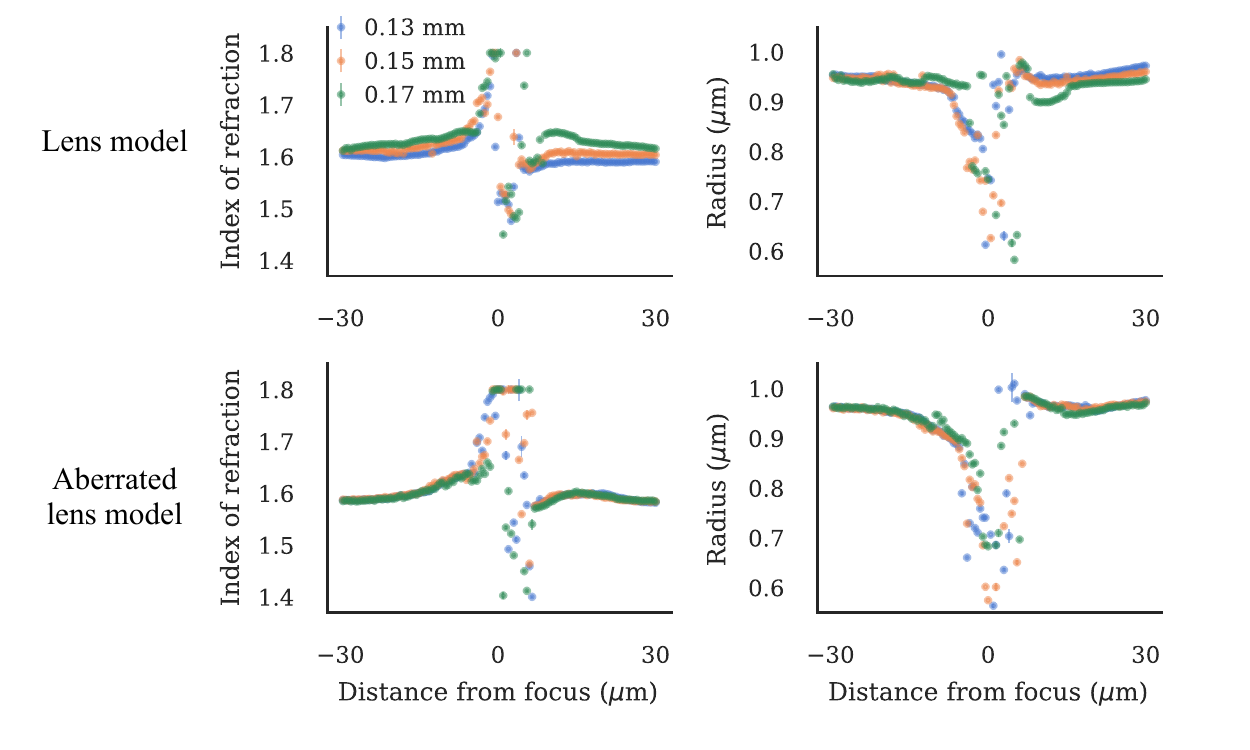}
  \caption{Refractive index and radius inferred by fitting the lens
    model and aberrated-lens models to the data, shown as a function of
    particle distance from the focal plane. Colors indicate the level of
    spherical aberration, as measured by the correction-collar setting
    (lowest level of aberration is at \SI{0.17}{mm}). Both models
    perform poorly near the focal plane, yielding unphysical and
    inconsistent values for the refractive index and radius when the
    particle is closer than about \SI{5}{\micro\meter} to the focus.
    Outside this region, the refractive index and radius inferred by
    fitting the lens model depend on the level of aberration in the
    system. For the aberrated-lens model, this systematic difference
    disappears, and we obtain consistent results across aberration
    levels.}
  \label{fig:params}
\end{figure}

Ignoring the effects of spherical aberration in the model results not
only in higher residuals but also in systematic shifts in the inferred
particle refractive index and radius with the aberration level. When the
particle is more than \SI{5}{\micro\meter} from the focal plane, the
inferred values for both the refractive index and radius differ
consistently between the most aberrated (\SI{0.13}{\milli\meter} collar
setting) and least aberrated (\SI{0.17}{\milli\meter}) systems, even
when the particle is at the same $\zp{}$ (Figure~\ref{fig:params}, top).
We quantify this parameter shift by determining the the absolute
difference between the inferred parameters for each $\zp{}$ more than
\SI{5}{\micro\meter} from the focal plane, then averaging across
particle position to find the mean absolute difference with standard
error. The inferred refractive indices differ by \SI{0.030(2)}{}, or
\SI{1.8(1)}{\percent}, and the inferred radii differ by
\SI{0.020(2)}{\micro\meter}, or \SI{2.0(2)}{\percent}, when the particle
is more than \SI{5}{\micro\meter} from the focal plane. This difference
is far larger than the uncertainty in parameters, estimated as the
standard error for each best-fit value and shown as error bars in
Figure~\ref{fig:params}.

When we account for aberration in our model, the inferred parameters no
longer depend on the level of aberration (Figure~\ref{fig:params},
bottom). When the sphere is more than \SI{5}{\micro\meter} from the
focal plane, the refractive indices inferred using the aberrated-lens
model differ by \SI{0.002(1)}{}, or \SI{0.16(3)}{\percent}, between the
most and least aberrated systems, an order-of-magnitude improvement over
the refractive indices inferred using the lens model. The inferred radii
differ by \SI{0.007(2)}{\micro\meter}, or \SI{0.8(2)}{\percent}, a
two-fold improvement.

Furthermore, the refractive index and radius that we infer by fitting
the aberrated-lens model are close to the values reported by the
manufacturer of the particles, irrespective of the aberration level.
When the microsphere is more than \SI{5}{\micro\meter} from the focal
plane, we infer a refractive index of $n=\SI{1.597(14)}{}$ and a radius
of $r=\SI{0.951(17)}{\micro\meter}$ for the least aberrated system, and
$n=\SI{1.599(15)}{}$ and $r=\SI{0.950(27)}{\micro\meter}$ for the most
aberrated system, where the values reported represent the mean and
standard deviation of inferred parameters across $\zp{}$ values. The
manufacturer reports the value of the radius as
$r=\SI{1.050(26)}{\micro\meter}$ and the refractive index as
$n=\SI{1.591}{}$ at \SI{590}{\nano\meter}; we expect the refractive
index of polystyrene to be $n=\SI{1.586}{}$ at
\SI{660}{\nano\meter}~\cite{ma_determination_2003}.  We find
good agreement between the inferred and expected refractive index
irrespective of aberration level in the system. Although there is a
difference between the manufacturer's mean value and our inferred
value of the radius, the values we infer are consistent across
aberration levels. The remaining systematic difference might arise
from differences in measurement conditions (dry \textit{versus}
aqueous), polydispersity in the particle stock, or additional
unmodeled optical effects. Nonetheless, the self-consistency of the
parameters and the agreement between the inferred and expected
refractive index show that modeling spherical aberration removes one
source of systematic error in particle characterization.

These characterization results do not change significantly with $\zp{}$
when the particle is more than \SI{10}{\micro\meter} from the focus. The
variation in the inferred radius, as quantified by the standard
deviation across $\zp$ values, is \SI{0.9}{\percent}, and that in the
inferred refractive index is \SI{0.5}{\percent}, as compared to
\SI{19.7}{\percent} (radius) and \SI{6.0}{\percent} (refractive index)
when the particle is less than \SI{10}{\micro\meter} from the focus.
Neither the lens model nor the aberrated-lens model give consistent or
even physically realistic estimates when the particle is within
\SI{5}{\micro\meter} of the focus, and in the
\SIrange{5}{10}{\micro\meter} range, the inferred values have a
noticeable $\zp{}$ dependence. For example, as the particle approaches
the focus from above, the refractive index is systematically
underestimated and the radius is systematically overestimated; the
reverse occurs as the particle approaches the focus from below.
Therefore, the forward-modeling approach that we demonstrate should be
used with caution when, for example, one must characterize particles
that could drift near the focal plane. Future work should focus on
modeling other optical effects that are relevant near the focus. We
hypothesize that the poor characterization close to the focus arises
because there are few fringes and little contrast, limiting the amount
of information that could be extracted by fitting. The dependence of the
inferred parameters on $z_p$ that we observe might provide some clues
about what other effects must be modeled to improve the precision near
the focal plane.

There is no additional computational cost to calculating a hologram with
the aberrated-lens model compared to the lens model. On a
\SI{1.6}{\giga\hertz} Intel Core i5 processor, the aberrated-lens model
takes $\SI{85(3)}{\milli\second}$ to generate a $200 \times 200$ pixel
hologram, while the lens model takes $\SI{86(5)}{\milli\second}$; these
numbers include overhead incurred by the \texttt{holopy}
package~\cite{barkley_holographic_2020}. However, it does take longer to
fit the aberrated-lens model to data because it has more parameters than
the lens model.

\section{Conclusion}
Unless carefully corrected, spherical aberration is present in typical
holographic microscopes and can significantly affect hologram structure.
Spherical aberration can arise in an otherwise corrected system when the
correction collar on a water-immersion lens is incorrectly set, or when
there are interfaces between media of different refractive indices---for
example, when an aqueous sample chamber is imaged using an oil-immersion
or air-immersion objective. We have shown that neglecting to account for
this aberration leads to inconsistent particle characterization when
fitting a forward model to the data.

Adding the effects of spherical aberration to a forward model of
hologram formation improves the fit of the model to the data and removes
aberration-dependent shifts in the recovered parameters. Fitting with
this aberrated-lens model makes particle characterization with
holography robust to aberration for both isolated spheres or
well-separated collections of spheres~\cite{fung2012imaging}. With
greater computational resources, it could be expanded to the
characterization of other particles, such as clusters and
spheroids~\cite{mackowski_calculation_1996, alexander_precise_2020}. The
robustness to level of aberration is a useful feature for experiments,
because it means that no prior knowledge or characterization of the
aberrations is needed. Instead, the aberration coefficients can be fit
at the same time as other parameters such as the refractive index and
radius. Thus, our method can correct not only the known aberrations in
the microscope, such as those induced by the coverslip, but also
aberrations that are unknown to the experimentalist because they are
intrinsic to the instrument or objective.

An interesting direction for future work is to determine whether
modeling the effects of aberration enables reliable particle
characterization even with a highly aberrated lens. For example,
low-cost holography with a ball lens may be possible. For highly
aberrated systems, our methodology could be extended to other types of
aberration, including curvature of field or coma.

\begin{appendices}

\section{Numerical methods}
\label{appendix:methods}

We evaluate Eqs.~\ref{eqn:integral_I0_def}--\ref{eqn:integral_I2_def}
using the same methods described by Leahy, Alexander, and
coworkers~\cite{leahy_large_2020} for an unaberrated lens. Because
spherical aberrations preserve the azimuthal symmetry in
Eqs.~\ref{eqn:integral_I0_def}--\ref{eqn:integral_I2_def}, their
numerical evaluation carries no additional computational complexity
relative to the unaberrated case and retains a computational advantage
over the lensless model due to additional numerical optimizations in the
lens model~\cite{leahy_large_2020}. We use the Python package
\texttt{holopy}~\cite{barkley_holographic_2020, holopy_github} to
calculate holograms using the lens model~\cite{leahy_large_2020}.

To fit these models, we use a multi-step approach designed to avoid
local minima. We first use a parallel-tempered, affine-invariant,
Markov-chain Monte Carlo ensemble sampler, as implemented by the Python
package \texttt{emcee}~\cite{foreman-mackey_emcee_2013}. We choose
uniform priors with bounds set by any physical constraints of the
experimental system ($x,y>0$, $r > 0$, $n>0$, $0<\beta<1.2$,
$\alpha>0$). We run this parallel-tempering algorithm at 7 temperatures
with 50 walkers for 2000 steps each, which takes approximately
\SI{12}{\hour} per hologram on one core. We therefore use parallel
tempering to fit only every 20th hologram in each stack.

We then fit the remaining intermediate holograms using the maximum
\textit{a posteriori} parameters found with parallel tempering as the
initial guess. We fit these holograms iteratively with the evolution
strategy CMA-ES (covariance matrix adaptation evolution
strategy)~\cite{hansen1996adapting}, as implemented by the Python
package \texttt{cma}~\cite{hansen_cma-espycma_2020}, and with a
Levenberg-Marquardt algorithm for nonlinear least-squares fitting, as
implemented by the python package
\texttt{lmfit}~\cite{newville_matthew_2014_11813}. We find that when we
fit with CMA-ES alone, we avoid local minima but often fail to fully
converge on the minimum within a reasonable computation time.
Conversely, we find that when we use only Levenberg-Marquardt
least-squares fitting, we converge to a good fit only with a very good
initial guess. Thus, we first explore the posterior landscape with
CMA-ES, and then ensure that we converge on the best fit through
least-squares minimization of the residuals. The combination of the two
algorithms with the initial guess from MCMC produces fits comparable to
those we find with parallel-tempering, but with much shorter run times.
When we maximize the posterior with CMA-ES, we choose broad Gaussian
priors, with means set by the maximum \textit{a posteriori} parameters
found by parallel tempering and variances set to physically reasonable
values, such as the width of a pixel for the particle's $x$ and $y$
coordinates ($\sigma_{x,y}=\SI{0.176}{\micro\meter}$,
$\sigma_z=\SI{4}{\micro\meter}$, $\sigma_n =0.2$,
$\sigma_r=\SI{0.1}{\micro\meter}$, $\sigma_\alpha=0.5$,
$\sigma_\beta=0.4$, $\sigma_{a_\ell} = 300$). The widest priors are
chosen for $a_0$, $a_1$, and $a_2$, which have widths of the same order
as the typical values we infer for the coefficients. We choose these
wide priors because in typical experiments the microscope user has
little \textit{a priori} knowledge of the aberration. For other
parameters, we place bounds on the Gaussian priors if there are any
physical constraints ($x,y>0$, $r > 0$, $n>0$, $0<\beta<1.2$,
$\alpha>0$). We then use least-squares fitting to minimize the residuals
between the data and the best-fit holograms found by CMA-ES. The error
bars shown in Figure~\ref{fig:params} are calculated by \texttt{lmfit}
from the estimated covariance matrix. Because the holograms are analyzed
individually, the uncertainties are independent. They do not account for
systematic errors. We find that when the holograms are closely spaced in
$\zp{}$, the parameters change slowly enough that we avoid local minima
in fitting the holograms between the select parallel-tempered frames.

To determine the order of the expansion of the aberration function
necessary to describe the data, we calculate the sum of squared
residuals $\chi^2$ and the maximum \textit{a posteriori} parameters for
increasing aberration order using the parallel-tempered sampler
described above. We select holograms taken at
$\zp{}=\SI{7.5}{\micro\meter}$ above the focal plane, for
\SIrange{0.13}{0.17}{\milli\meter} correction. Holograms at this
position are well within the distance from the focus where fitting
returns reasonable residuals for all models, but they also show a large
change in structure as we change the level of aberrations, indicating a
strong dependence on the aberration function. We find that the $\chi^2$
value decays with increasing aberration expansion order, with $\chi^2$
at a maximum with zeroth-order aberration (lens model) and decreasing
with each additional aberration order included before plateauing. We
observe this decay in $\chi^2$ for all levels of aberration, with the
sharpest decay occurring in the most aberrated system. For our data, the
$\chi^2$ values plateau when we expand up to the eighth-order phase
polynomial, which includes aberration coefficients $a_0$, $a_1$, 
and $a_2$. We find very small differences between the eighth-order
and tenth-order expansion, with an average decrease in $\chi^2$ of
\SI{0.6}{\percent} across levels of aberration. This improvement in
goodness-of-fit is much smaller than the \SI{43}{\percent} mean decrease
in $\chi^2$ found between the zeroth- and eighth-order expansion.

We see additional evidence favoring the eighth-order expansion in the
comparison of the inferred maximum \textit{a posteriori} parameters to
the manufacturer's specifications for the particles. We find that when
we include only $a_0$ in the expansion, the $\chi^2$ value decreases
greatly compared to the lens-model, but the maximum \textit{a
  posteriori} parameters become unphysical, with the maximum \textit{a
  posteriori} refractive index jumping to nearly 1.68, much larger than
the anticipated value of 1.59. As we increase the number of aberration
coefficients, the maximum \textit{a posteriori} values of the parameters
plateau to physical values with the eighth-order expansion with
coefficients $a_0$, $a_1$, and $a_2$. We find small differences between
characterization with the eighth- and tenth-order expansion, with
average differences of \SI{0.5}{\percent} in refractive index and
\SI{0.6}{\percent} in radius. To avoid unnecessary complexity in the
model, we choose to expand the aberration function to the eighth-order
phase polynomial and include three aberration parameters in the model.

Although we find that truncating the aberration series at the
eighth-order phase polynomial is appropriate for the aberration level in
our experimental system, other experimental systems may require
higher-order expansions to fully capture the aberration function.

\end{appendices}

  \section*{Funding}
  This work is partially supported by the National Science Foundation
  through the Harvard University Materials Research Science and
  Engineering Center under grant numbers DMR-1420570 and DMR-2011754.
  Additional support was provided by the Department of Defense through
  the National Defense Science and Engineering Graduate Fellowship
  (NDSEG).

  \section*{Acknowledgments} We thank R. Alexander and S. Barkley for
  useful discussions on fitting methods.

  \section*{Disclosures}
  
  \noindent The authors declare no conflicts of interest.

  \section*{Data availability} Data from the experiments shown in this
  paper are available in Ref.~\citenum{martin21:_data_effec}. Source
  code for the forward model and inference calculations is available in
  Ref.~\citenum{holopy_github}.
  

\bibliography{sphere_ab_bib}

\end{document}